\documentclass{aa}
\usepackage{txfonts}
\usepackage{graphicx}

\begin{document}
 
\title{Constraining the Cosmological Density of Compact Objects with the Long-Term Variability of Quasars} 
\markboth{E. Zackrisson \& N.Bergvall: }{Constraining Dark Matter with the Long-Term Variability of Quasars}
\authorrunning{E. Zackrisson \& N. Bergvall}
\titlerunning{Constraining Dark Matter with the Long-Term Variability of Quasars} 
\author{Erik Zackrisson \and Nils Bergvall}

\offprints{Erik Zackrisson, 
\email{Erik.Zackrisson@astro.uu.se}}

\institute{Astronomiska observatoriet, Box 515, S-75120 Uppsala, Sweden}

\date{Received 0000 / accepted 0000}

\abstract{By comparing the results from numerical microlensing simulations to the observed long-term variability of quasars, strong upper limits on the cosmological density of compact objects in the $10^{-4}\ M_\odot$--$1 \ M_\odot$ range may in principle be imposed. Here, this method is generalized from the Einstein-de Sitter universe to the currently favored $\Omega_\mathrm{M}=0.3$, $\Omega_\Lambda=0.7$ cosmology and applied to the latest observational samples. We show that the use of high-redshift quasars from variability-selected samples has the potential to substantially improve current constraints on compact objects in this mass range. We also investigate to what extent the upper limits on such hypothetical dark matter populations are affected by assumptions concerning the size of the optical continuum-emitting region of quasars and the velocity dispersion of compact objects. We find that mainly due to uncertainties in the typical value of the source size, cosmologically significant populations of compact objects cannot safely be ruled out with this method at the present time.\keywords{Dark matter -- gravitational lensing -- quasars: general -- cosmology: miscellaneous}}

\maketitle

\section{Introduction}
Despite much effort to detect or rule out cosmologically significant populations of compact bodies in the stellar to planetary mass range, such objects still remain viable candidates for the dark matter of the universe. Although many of the compact dark matter candidates proposed are baryonic (e.g. red, white and brown dwarfs, neutron stars, stellar black holes, gas clouds) and therefore constrained by standard Big Bang nucleosynthesis and the determination of the primeval deuterium abundance to contribute no more than $\Omega_\mathrm{B}h_{100}^2=0.02$ (e.g. Burles et al. \cite{Burles et al.}) to the cosmological density, several candidates do circumvent these constraints, either by being non-baryonic or by making their baryonic content unavailable by the time of nucleosynthesis: e.g. primordial black holes (Hawking \cite{Hawking}), quark nuggets (Alam et al. \cite{Alam et al.}), mirror matter MACHOs (Mohapatra \cite{Mohapatra}) and aggregates of bosons or fermions (Membrado \cite{Membrado}). A large number of methods do however exist to constrain the cosmological densities of such populations (see Dalcanton et al. \cite{Dalcanton et al.} and Carr \& Sakellariadou \cite{Carr & Sakellariadou} for reviews). 

In this paper, we will be concerned with compact objects in the potentially very interesting mass range $10^{-4}\ M_\odot$--$1 \ M_\odot$, where indirect detections of cosmologically significant populations have been suggested (e.g. Hawkins \cite{Hawkins1996}). In this region, the currently most powerful constraints on the cosmological density of such objects (regardless of type), $\Omega_{\mathrm{compact}}$, come from Dalcanton et al. (\cite{Dalcanton et al.}) and Schneider (\cite{Schneider}, hereafter S93), and are both based on theoretically predicted effects of quasar microlensing. 

By comparing the equivalent width distribution of quasar emission lines predicted by microlensing scenarios to that of observed samples, Dalcanton concludes $\Omega_{\mathrm{compact}}\le 0.2$ for $10^{-3}\ M_\odot$--$1 \ M_\odot$ compact objects in a critical universe. Taken at face value, the constraints from S93 are however even stronger at the lowest masses: $\Omega_{\mathrm{compact}}\le 0.1$ for $10^{-3}\ M_\odot$--$3\cdot10^{-2}\ M_\odot$. 

The method of S93 is based on the argument that large populations of compact objects should statistically induce variations in quasar light curves larger than those actually observed. The limits derived this way do however rely on the premise that the many parameters going into the microlensing simulations can be sufficiently well constrained by observations or reasonable assumptions, effectively making $\Omega_{\mathrm{compact}}$ the only free parameter. The aim of this paper is to investigate whether this may actually be accomplished at the present time. Of particular importance is the ill-determined typical size of the UV--optical continuum-emitting region of quasars, $R_{\mathrm{QSO}}$, which S93 assumes to be $R_{\mathrm{QSO}}=10^{13}$ m when deriving the constraints quoted above. 

Here, the method of S93 will be generalized from the Einstein-de Sitter (EdS) universe to the currently favored $\Omega_\mathrm{M}=0.3$, $\Omega_\Lambda=0.7$ cosmology and applied to the latest observational samples (Hawkins \cite{Hawkins2000}, hereafter H2000). The sensitivity of the constraints to uncertainties in $R_{\mathrm{QSO}}$ and the velocity dispersion of compact objects will be evaluated using recently developed methods to better approximate the magnification in the case of large source microlensing.

\section{Numerical simulations of quasar microlensing}
\subsection{Computational method}
The method used to derive the statistical properties of light curves of quasars microlensed by a cosmological distribution of compact objects is based on the machinery outlined in S93, extended to arbitrary Friedmann-Lema\^{\i}tre cosmologies using the angular size distances of Kayser et al. (\cite{Kayser et al.}) and to the case of large-source microlensing using the magnification formula of Surpi et al. (\cite{Surpi et al.}).

In this technique, the multiplicative magnification approximation (Ostriker \& Vietri \cite{Ostriker & Vietri}) is assumed to adequately reproduce the statistical probability of variability. In this case, the magnification, $\mu_\mathrm{tot}$, due to $i$ microlenses is equal to the product of the individual ones:
\begin{equation} 
\mu_\mathrm{tot}=\prod\limits_{i}\mu_i.
\label{eq1}
\end{equation}
Although this approximation cannot be expected to reproduce detailed features of microlensing light curves, it has been proved useful (e.g. Pei \cite{Pei 1993}) for statistical investigations and significantly reduces the calculation time required by ray-tracing and similar methods. 

We follow conventions commonly found in gravitational lensing literature and  use $D_\mathrm{d}$, $D_\mathrm{s}$ and $D_\mathrm{ds}$ to denote the angular size distances from observer to lens, observer to source and from lens to source, respectively. If $\xi$ is the separation of the lens to the line that joins the source and observer, the dimensionless impact parameter y (in units of the Einstein radius of the lens) then becomes
\begin{equation} 
y=c\xi\sqrt{D_\mathrm{s}/(4GM_\mathrm{compact}D_\mathrm{d}D_\mathrm{ds})} 
\label{eq2}
\end{equation}
where $M_\mathrm{compact}$ is the mass of the compact object acting as a microlens. The dimensionless source radius R may similarly be expressed as
\begin{equation}
R=cR_\mathrm{QSO}\sqrt{D_\mathrm{d}/(4GM_\mathrm{compact}D_\mathrm{s}D_\mathrm{ds})}.
\label{eq3}
\end{equation}

For a circular source with uniform brightness, the magnification due to a single microlens is in the limit of small sources ($R < 1.2$) approximated by (Schneider \cite{Schneider}):
\begin{eqnarray}
\mu(y,R)\approx\nonumber \frac{2(R^2+2)}{R(R^2+4)^{3/2}}\left[ 1+\frac{(R^2+2)y^2}{R^2(R^2+3)} \right]^{-2} + \\ \left\{ \begin{array}{ll} \frac{(R^2+3)^2+3-4(y/R)^2}{R(4+R^2)^{3/2}} & \mbox{for } 
 y\le R \\ \frac{2+y^2}{y\sqrt{y^2+4}} & \mbox{for } y>R \end{array}
 \right. .
\label{eq4}
\end{eqnarray}
In the limit of large sources ($R \ge 1.2$) the following approximation (Surpi et al. \cite{Surpi et al.}) is instead used:
\begin{eqnarray}
\log(\mu(y,R))\approx \nonumber \log(\sqrt{1+4/R^2})\cdot  \hspace{3cm} \\  \hspace{2cm} \left\{ \begin{array}{ll} 1 & \mbox{for } 
 y<R-1 \\ \frac{1}{2}(1-y+R) & \mbox{for } R-1<y<R+1 \\ 0 & \mbox{for } y>R+1 \end{array} \right. .
\label{eq5}
\end{eqnarray}
 The redshift interval from the observer to the source is divided into lens planes which are randomly populated by microlenses with masses and velocities following some arbitrary distribution. All lens planes are evenly distributed in redshift between the source and observer, with a linear separation given by (distance by light travel time):
\begin{equation}
D^{\mathrm{c}}_{\Delta z}=\frac{c}{H_0}\int\limits_{\Delta z}  \frac{\mathrm{d}z}{(1+z)\sqrt{Q(z)}}
\label{eq6}
\end{equation}
where
\begin{equation}
Q(z)=\Omega_\mathrm{M}(1+z)^3-(\Omega_\mathrm{M}+\Omega_\Lambda-1)(1+z)^2+\Omega_\Lambda .
\label{eq7}
\end{equation}
Due to circular symmetry, it is sufficient to consider rectangular lens planes centered on the source with lenses moving in one dimension only. If we assume the lenses to move along the x-axis, there are two relevant boundaries $l_y$ of the lens plane in the perpendicular direction, depending on whether the source is small or large. If $R_\mathrm{E}$ represents the Einstein radius of the lenses, 
\begin{eqnarray}
l_y= \left\{\begin{array}{ll} 2 R_\mathrm{E} y_\mathrm{max} & \mbox{for } R<1.2  \\ 2(R_\mathrm{QSO}(D_\mathrm{d}/D_\mathrm{s}) + R_\mathrm{E}) & \mbox{for } R\ge 1.2\\ \end{array} \right. ,
\label{eq8}
\end{eqnarray}
where $l_y$ in the case of $R\ge1.2$ corresponds to twice the size of the boundary of the outer non-zero magnification zone in (\ref{eq5}). In the case of $R<1.2$, S93 shows that $y_\mathrm{max}=15$ is sufficient to ensure that the contribution to the magnification from lenses outside the considered region is negligible.

In the case of an extended distribution of lens masses, $l_y$ should be set equal to whichever of the two expressions in (\ref{eq8}) that produces the highest value when evaluated for the maximum $R_\mathrm{E}$ of any mass considered. 

The extent of the lens plane along the x-axis is derived by adding $l_y$ to a term which ensures that no lens originally outside the lens plane will move sufficiently close to the source during the time span $\Delta t$ of the simulation to give a non-negligible contribution to the magnification (see S93):
\begin{equation}
l_x=l_y+2\Delta t(v_\mathrm{obs}+\max |v_\mathrm{compact}|)
\label{eq9},
\end{equation}
where $z_\mathrm{d}$ is the redshift of the lens plane and $v_\mathrm{obs}$ and $v_\mathrm{compact}$ are the observer and lens velocities, respectively, perpendicular to the line-of-sight. In the case of a Gaussian distribution of lens velocities, $\max |v_\mathrm{compact}|$ may be taken to be the $3\sigma$-deviation. 

For the lens population, a constant comoving density is assumed. In the case where all compact objects share the same mass, the number of lenses populating a lens plane becomes:
\begin{equation}
N_\mathrm{compact}=\frac{\rho_\mathrm{crit,0}(1+z_\mathrm{d})^3 \Omega_\mathrm{compact} D^c_{\Delta z}l_x l_y} {M_\mathrm{compact}}
\label{eq10}.
\end{equation}
At each time step during the simulation, the observer, source and lenses are relocated according to their individual velocities and the total magnification calculated using (\ref{eq1}). The resulting time-dependent magnification $\mu_\mathrm{tot}(t)$ represents the simulated light curve. S93 may be consulted for further details on this method. 

\subsection{General assumptions}
In this paper, a $\Lambda$-dominated cosmology will be assumed, in which $\Omega_\mathrm{M}=0.3$, $\Omega_\Lambda=0.7$ and H$_0=65$ km/s/Mpc. The mean density inside the simulated volume is assumed to be equal to the mean density of the universe, which implies a homogeneity parameter (Kayser et al. \cite{Kayser et al.}) $\eta=1.0$. All compact objects are furthermore considered to be randomly distributed. 

When demonstrating how the constraints on $\Omega_{\mathrm{compact}}$ depend on different parameters, we will assume all compact lenses to have the same mass $M_\mathrm{compact}$. Unless stated otherwise, the velocities of compact objects perpendicular to the line-of-sight are assumed to be normally distributed with 2D velocity dispersions $\sigma_{\mathrm{v,compact}}=400$ km/s. 

Since the quasars used in this investigation are all located in the direction $\alpha =21^\mathrm{h} 28 ^\mathrm{m}$, $\delta=-45\degr$, a universal observer velocity will be used. Adopting the velocity of the Sun relative to the cosmic microwave background derived by Lineweaver et al. (\cite{Lineweaver et al.}), the velocity of the observer perpendicular to the line-of-sight becomes $307.9$ km/s. 

\subsection{Tests}
\begin{figure}[t]
\resizebox{\hsize}{!}{\includegraphics{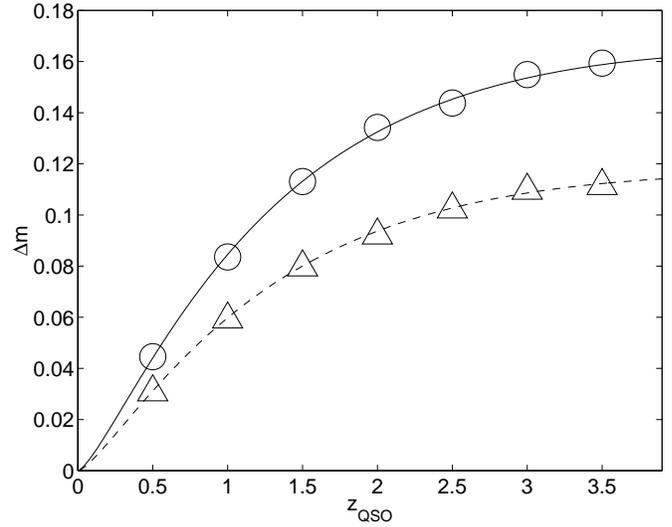}}
\caption[]{Analytical $\Delta m$ (Surpi et al. \cite{Surpi et al.}) compared to $\Delta m$ derived from simulated light curves. At each quasar redshift, $z_\mathrm{QSO}$, the $\Delta m$ derived from simulations has been based on 50 generated light curves, each spanning 2000 years and sampled at intervals of 1 year. Lines represent the analytical predictions for a $\Lambda$-dominated universe ($\Omega_{\mathrm{M}}=0.3$, $\Omega_\Lambda=0.7$) with $\Omega_{\mathrm{compact}}=0.3$ (solid) and $\Omega_{\mathrm{compact}}=0.15$ (dashed). In both cases, we have assumed $R_\mathrm{QSO}=3\cdot 10^{13}$ m. Circles and triangles indicate the corresponding results from simulations.}
\label{deltamtest}
\end{figure}

The implementation of the algorithm has been subjected to several tests. It passes the test concerning the mean number of lenses with impact parameter $y \le 1$ as outlined in S93. When the magnification formulas (\ref{eq4} \& \ref{eq5}) are modified to accord with the definition of magnification peaks used in Alexander (\cite{Alexander}), the characteristic time scale of microlensing events as a function of lens redshift also agrees well with the analytical expressions in that paper. Finally, the standard deviation in magnitudes $\Delta m$ around the mean magnification of the simulated light curves, calculated as
\begin{equation} 
\Delta m=\sqrt{\overline{m(t)^2}-\overline{m(t)}^2}, 
\label{eq11}
\end{equation}
where the deviation in magnitudes $m(t)$ from the mean is given by
\begin{equation} 
m(t)=2.5\log{\mu_\mathrm{tot}(t)}-\overline{2.5\log \mu_\mathrm{tot}(t)}, 
\label{eq12}
\end{equation}
almost perfectly matches the analytical expression derived for large sources in Surpi et al. (\cite{Surpi et al.}), as demonstrated in Fig.~\ref{deltamtest}. 

For the current investigation, 30 lens planes have been used. The adequacy of this number has been verified by increasing the number of lens planes by a factor of two without any significant impact on the results.

\subsection{Flux conservation}
Magnification, as used here, is defined relative to the smooth Friedmann-Lema\^{\i}tre universe, which is not a consistent treatment in terms of flux conservation. Even though this inconsistency has no effect on the variability of simulated light curves, the properly normalized scale of magnification becomes important when simulating the effects of amplification bias. To force $\overline{\mu_\mathrm{tot}}=1$ when taking the average along all lines of sight to a source, we follow a recipe developed by Canizares (\cite{Canizares}) in which the corrected magnification $\mu_\mathrm{tot}^\mathrm{corr}(t)$ is calculated from the uncorrected $\mu_\mathrm{tot}^\mathrm{uncorr}(t)$ using
\begin{equation} 
\mu_\mathrm{tot}^\mathrm{corr}(t)=f_\mathrm{corr}\cdot\mu_\mathrm{tot}^\mathrm{uncorr}(t). 
\label{eq13}
\end{equation} 
Here, $f_\mathrm{corr}$ is a correction factor given by 
\begin{equation} 
f_\mathrm{corr}=1/\overline{\mu_\mathrm{tot}^\mathrm{uncorr}(t)}
\label{eq14}
\end{equation} 
where the average $\overline{\mu_\mathrm{tot}^\mathrm{uncorr}(t)}$ is taken over all light curves generated with randomized lens positions but otherwise identical input parameters.   

\section{How to infer upper limits on the cosmological density of compact objects from the long-term variability of quasars}

\subsection{General method}
Even though the observed optical variability of quasars on time scales of a few years (e.g. Hook et al. \cite{Hook et al.}, V\'eron \& Hawkins \cite{Véron & Hawkins}, Cristiani et al. \cite{Cristiani et al.}) could be a combined effect of intrinsic (e.g. accretion disk instabilities, supernova explosions) and extrinsic (e.g. microlensing) variations, combining the two proposed mechanisms can however only increase the probability of flux variations. By assuming that {\it all} variability is due to microlensing and comparing the predictions from microlensing scenarios to the observed variation probabilities of quasar light curves, upper limits on the cosmological density of compact object may therefore be imposed. This technique was first implemented in S93 to constrain compact dark matter populations in the mass range $10^{-4}\ M_\odot$--$1 \ M_\odot$ for an EdS universe, using the observational sample of Hawkins \& V\'eron (\cite{Hawkins & Véron}, hereafter HV93).

The method for constraining $\Omega_{\mathrm{compact}}$ outlined in S93 is based on the amplitudes, $\delta m$, of quasar light curves. Here, the amplitude is defined as the difference between the minimum and maximum yearly magnitudes observed in a quasar within the duration of the monitoring programme. Each yearly magnitude is furthermore the average of roughly four intrayear measurements. For the case of variations induced purely by microlensing, this reduces to:
\begin{equation} 
\delta m=\max (2.5\log  \mu_{\mathrm{tot}}(t))-\min (2.5\log  \mu_{\mathrm{tot}}(t)), 
\label{eq15}
\end{equation}
where the yearly magnitudes $2.5\log  \mu_{\mathrm{tot}}(t)$ for each integer year $t$ are formed from four evenly spaced intrayear magnification data points.

When comparing observed and synthetic quasar samples, a simple statistical analysis is used. The cumulative probability $P(\delta m)$ predicted by microlensing simulations of finding amplitudes higher than $\delta m$ is used to calculate the expected number of objects in an observed sample of size $N_\mathrm{sample}$ with amplitudes higher than $\delta m$:
\begin{equation}
<N(\delta m)> = N_\mathrm{sample} P(\delta m).
\label{eq16}
\end{equation}
The Poisson probability, $P(N(\delta m))$, that the number of objects in the synthetic sample with amplitudes higher than $\delta m$ is smaller than, or equal to the observed number $N(\delta m)$ then becomes 
\begin{equation}
P(N(\delta m))=e^{-<N(\delta m)>}\sum_{n=0}^{N(\delta m)} \frac{<N(\delta m)>^n}{n!}.
\label{eq17}
\end{equation}
The minimum of this probability
\begin{equation}
P_\mathrm{min}=\min_{\delta m} P(N(\delta m)),
\label{eq18}
\end{equation}
essentially corresponds to the probability that the missing fraction of high amplitudes in the observed sample is simply due to the finite number of observed objects. Here, it will be used as a measure of the probability that the observed sample is compatible with a particular microlensing scenario. 

In principle, the statistical properties of synthetic samples depend on a large number of parameters (for the background cosmology $\Omega_\mathrm{M}$, $\Omega_\Lambda$, H$_0$ and $\eta$; for the lens population $\Omega_\mathrm{compact}$, $\sigma_\mathrm{v, compact}$ and parameters describing the mass spectrum of compact objects; for the quasar population $\sigma_\mathrm{v, QSO}$ and the distribution of $z_\mathrm{QSO}$ in addition to parameters describing the luminosity function and the $L_\mathrm{QSO}$-$R_\mathrm{QSO}$ relation; for the observer the velocity perpendicular to the line-of-sight). If most of these can be constrained by measurements or reasonable assumptions, upper limits on $\Omega_{\mathrm{compact}}$ may be imposed. In order to facilitate the visualization of the constraints imposed by this method, and still enable a direct comparison to the results of S93, a threshold value of $P_\mathrm{min}=10\%$ has been adopted to delimit allowed and rejected regions of the microlensing parameter space. When $P_\mathrm{min}$ falls below this value, the microlensing scenario responsible is considered to be inconsistent with the observations and can be ruled out. A limit of $P_\mathrm{min}=10\%$ corresponds to the case where the microlensing scenario has predicted the presence of 2.3 quasars with amplitudes higher than some arbitrary value in a sample, yet none is observed.

\subsection{Transition to the $\Lambda$-dominated universe and the use of the large-source amplification formula}
What impact does the transition from an EdS ($\Omega_\mathrm{M}=1.0$, $\Omega_\Lambda=0.0$) to a $\Lambda$-dominated universe ($\Omega_\mathrm{M}=0.3$, $\Omega_\Lambda=0.7$) have on the constraints derived? Since the  $\Lambda$-dominated cosmology implies a higher angular size distance out to a particular redshift, more lenses will contribute to the light curve for a particular value of $\Omega_\mathrm{compact}$. In Zackrisson et al. (\cite{Zackrisson et al.}) we showed that the constraints on $\Omega_\mathrm{compact}$ were significantly strengthened due to this effect. 

Since the approximation (\ref{eq4}) used in S93 underestimates the magnification in the limit of large sources, Surpi et al. (\cite{Surpi et al.}) predicted that the implementation of (\ref{eq5}) should improve constraints of S93 type. In Zackrisson et al. (\cite{Zackrisson et al.}), it was however shown that in the range $M_\mathrm{compact}=10^{-4} \ M_\odot$--$1 \ M_\odot$ this correction only had a modest impact on the upper limits on $\Omega_{\mathrm{compact}}$ in a part of parameter space ($M_\mathrm{compact}=10^{-4} M_\odot$, $R_\mathrm{QSO} \ge 10^{13}$ m) for which no interesting constraints could be inferred. 

As becomes evident when comparing the constraints from Zackrisson et al. (\cite{Zackrisson et al.}) to those of S93, there exists a systematic difference between the cumulative probabilities $P(\delta m)$ derived when analyzing the samples of HV93, even when the simulations are matched as closely as possible given the published information. It seems that the model of S93 systematically predicts somewhat higher probabilities of large amplitudes. For this reason, the constraints derived for the $\Lambda$-dominated cosmology are actually only slightly stronger than those originally presented in S93. The origin of this discrepancy is not known, but could be due to the redshift resolution or the value of H$_0$ used.

\subsection{Uncertainties in $R_{\mathrm{QSO}}$}
The microlensing magnification of a distant source is highly sensitive to the source size assumed. Even though the maximum magnification from a single lens decreases with increasing source size, more lenses may at the same time contribute to the microlensing effect. As shown in Surpi et al. (\cite{Surpi et al.}), the variability amplitude does however drop as $R_{\mathrm{QSO}}$ is increased for a fixed lens mass within the parameter space explored.

The typical value of the parameter $R_{\mathrm{QSO}}$ is far from well-determined, but certain constraints do exist. If we assume the UV--optical continuum of quasars to originate in the accretion disc surrounding the central black hole, we may expect $R_{\mathrm{QSO}}$ to be at least a few times larger than the gravitational radius of the central mass, which for an $10^8 M_\odot$ objects amounts to $1.5 \cdot 10^{11}$ m.  

Tight upper limits on  $R_{\mathrm{QSO}}$ are also available for a few well-studied quasars for which measurements of duration or magnification of microlensing events have been made. For the gravitational lens system QSO 0957+561, Refsdal et al. (\cite{Refsdal et al.}) have been able to place the constraint  $R_{\mathrm{QSO}}<6\cdot 10^{13}$ m at a significance level of 10\%. For the same system Pelt et al. (\cite{Pelt et al.}) have argued that $R_{\mathrm{QSO}}\approx 3\cdot 10^{13}$ m (within a factor of 3). Shalyapin (\cite{Shalyapin}) has similarly inferred $R_{\mathrm{QSO}}<1\cdot 10^{13}$ m for QSO 2237+0305, with a most likely value at $R_{\mathrm{QSO}}\approx 6\cdot 10^{12}$ m. It is however not clear how typical these values are of the quasar population as a whole. 

The constraints on $\Omega_\mathrm{compact}$ quoted from S93 are based on the assumption that $R_{\mathrm{QSO}}=10^{13}$ m. In the following, we will however investigate to what extent the constraints on $\Omega_{\mathrm{compact}}$ are affected when $R_{\mathrm{QSO}}$ is allowed to take on higher and lower values inside the range $R_{\mathrm{QSO}}=10^{12}$--$10^{14}$ m, which is an interval commonly considered in studies of quasar microlensing (e.g. Tadros et al. \cite{Tadros et al.}).

\subsection{Building synthetic samples} 
In addition to the parameters that determine the variability of an individual quasar, distributions of $v_\mathrm{QSO}$, $z_\mathrm{QSO}$, $R_\mathrm{QSO}$ and $L_\mathrm{QSO}$ also need to be considered when building the synthetic quasar samples used for comparison with the observed amplitude distributions. 

\subsubsection{The velocity distribution of quasars}
Even though the clustering properties of quasars as a function of redshift are poorly determined, the tendency to cluster does not appear sufficiently strong to jeopardize a comparison with the kinematics of galaxies. We will therefore use galaxy data to estimate the velocity dispersion $\sigma_\mathrm{v,QSO}$ of quasars perpendicular to the line-of-sight. 

In an investigation of synthetic galaxy catalogues from the Virgo consortium (Coil et al. \cite {Coil}) one finds a typical line-of-sight velocity dispersion of galaxies within a 1 Mpc box at $z=1$ of $\sigma_\mathrm{v} \approx 180\pm20$  km/s. If large scale motions are included, the value increases to roughly 300 km/s. Empirical results from the Las Campanas survey (Baker et al. \cite {Baker}), corresponding to low $z$, result in slightly lower values. Based on these numbers, and taking into account the fact that quasars are predominantly located at high $z$, we will adopt a universal quasar velocity dispersion of $\sigma_\mathrm{v,QSO} = 300$ km/s.

\subsubsection{Amplification bias}
A quasar that lacks sufficient intrinsic brightness to be included in a flux-limited sample may still be magnified through gravitational microlensing to reach above the threshold for detection. A flux-limited sample will therefore contain an enhanced fraction of highly magnified objects. The correlation between minimum magnification and amplitude pointed out by S93 implies that a flux-limited sample should also display an enhanced probability for large variations. When deriving $P(\delta m)$ for a synthetic sample, the effects of amplification bias should therefore be taken into account.

In order to simulate the impact of amplification bias on the flux-limited samples of HV93 and H2000, we have assumed the optical quasar luminosity function (LF) derived by Boyle et al. (\cite{Boyle et al.}) for a $\Omega_\mathrm{M}=0.3$,  $\Omega_\Lambda=0.7$ cosmology:  
\begin{equation} 
\Phi(M_\mathrm{B},z) = \frac{\Phi^*}{10^{0.4[(\alpha+1)(M_\mathrm{B}-M_\mathrm{B}^*(z))]}+10^{0.4[(\beta+1)(M_\mathrm{B}-M_\mathrm{B}^*(z))]}}
\label{eq19}
\end{equation}
where the redshift evolution is given by 
\begin{equation}
M_\mathrm{B}^*(z)=M_\mathrm{B}^*(0)-2.5(k_1z+k_2z^2)
\label{eq20}
\end{equation}
and $\alpha=-3.41$, $\beta=-1.58$, $M_\mathrm{B}^*(0)=-22.65$, $k_1=1.36$, $k_2=-0.27$, $\Phi^*=0.36\cdot 10^{-6}$ $\mathrm{Mpc}^{-3}\mathrm{mag}^{-1}$.

Using this LF, the probability for a quasar to have a certain $M_\mathrm{B}$ as a function of $z$ may be derived. 

When building a synthetic sample subject to amplification bias, a large number of light curves is first generated with uniform $z_\mathrm{QSO}$ distribution within the redshift span of the observational sample. For each of these, an absolute magnitude  following the derived $M_\mathrm{B}$ probability is then randomly generated. These $M_\mathrm{B}$ are converted into intrinsic apparent magnitudes $m_\mathrm{B,0}$ assuming a power-law continuum with slope $\alpha=0.5$ for all quasars when calculating the k-correction. The effect of emission lines and deviations from a simple power-law continuum on the k-correction has been neglected, since at least for $0.4<z_\mathrm{QSO}<2.2$ objects the $\alpha=0.5$ k-correction closely resembles the average k-correction derived from observations (Wisotzki \cite{Wisotzki}). 

Since the observed quasar magnitudes are defined differently in the HV93 and H2000 samples, the $m_\mathrm{B,0}$ associated with each synthetic light curve is corrected for the effects of microlensing using different formulas depending on the observational sample used for comparison:
\begin{equation} m_\mathrm{B}=m_\mathrm{B,0}-\left\{ \begin{array}{ll} \min(2.5 \log({\mu_\mathrm{tot}(t)})) & \mbox{for HV93} \\ 2.5 \log(\mu_\mathrm{tot}(0)) & \mbox{for H2000}\end{array} \right. 
\label{eq21}
\end{equation}
where $2.5\log({\mu_\mathrm{tot}(t)})$ are the yearly magnitudes defined in section 3.1. All objects fainter than the $m_\mathrm{B}$ threshold used to define the observational sample are then rejected for efficiency reasons. This procedure is repeated until a sufficiently large number of synthetic light curves have passed the threshold. Finally, the actual $m_\mathrm{B}$ and $z_\mathrm{QSO}$ distributions of the observed samples are reproduced by dividing both synthetic and observational samples into identical bins and for each bin randomly selecting a number of objects from the synthetic sample proportional to the relative content of the corresponding observational bin.

This procedure of generating synthetic samples with simulated amplification bias differs from the one adopted by S93 in the assumed LF, the enforced flux conservation and the more realistic $m_\mathrm{B}$ and $z_\mathrm{QSO}$ distributions of the final samples. A comparison between $P_\mathrm{min}$ derived from synthetic samples with and without enforced observational flux distributions shows that the amplification bias produced by this method turns out to be quite modest, and most closely resembles the $s=0$ (weakest bias considered) scenario explored in S93. 

In simulating the amplification bias, we have assumed the observed LF to be identical to the intrinsic one, thereby neglecting the effects that lensing could have on the observed LF itself. Since the effect of lensing by compact objects as well as isothermal galaxies and clusters is to decrease the proportion of low-luminosity quasars compared to the intrinsic LF (e.g. Pei \cite{Pei 1995}), we are in fact underestimating the amplification bias by adopting the observed LF. This ensures that the upper limits on $\Omega_\mathrm{compact}$ derived are conservative. 

\subsubsection{The $L_\mathrm{QSO}$-$R_\mathrm{QSO}$ Relation}
In relating the intrinsic luminosity of a quasar to the size of the UV--optical continuum-emitting region, two different models have been considered. 

First, we have considered one in which $R_\mathrm{QSO}$ is assumed unrelated to $L_\mathrm{QSO}$ and treated as a free parameter in the range $10^{12}$--$10^{14}$ m. 

We have also experimented with the thin accretion disk relation of Czerny et al. (\cite{Czerny et al.}) which, assuming a power-law continuum with index $\alpha=0.5$, may be written (SI units): 
\begin{equation}
R_\mathrm{QSO}=3.37\cdot 10^{20}\frac{\lambda_\mathrm{eff}^{5/2} L_{\lambda_\mathrm{eff}}^{1/2}}{c^2(1+z)^{7/4}}.
\label{eq22}
\end{equation}
In this study, the effective wavelength $\lambda_\mathrm{eff}$ has been taken to be 4400 \AA $\;$since we are dealing with measurements in the B-band. Assuming Johnson filters and $\alpha=0.5$, $L_{\lambda_\mathrm{eff}}$ (J/s/m) is approximately related to $M_\mathrm{B}$ through
\begin{equation}
L_{\lambda_\mathrm{eff}}=6.77\cdot 10^{34}\cdot 10^{M_\mathrm{B}/-2.5}.
\label{eq23}
\end{equation}
When generating the synthetic samples in the case where $L_\mathrm{QSO}$ is assumed related to  $R_\mathrm{QSO}$, (\ref{eq22}) is used in conjunction with (\ref{eq23}) to assign each quasar with a certain intrinsic $M_\mathrm{B}$ to one of the light curves with the closest discrete value of $R_\mathrm{QSO}$ from the grid described in section 3.4.4. 

\subsubsection{Parameter values and size of synthetic samples} 
When generating the grid of synthetic samples used in the comparison with the observations, all combinations of the discrete parameter values listed in Table 1 have been used, giving a total of 150 simulation configurations. To ensure that the magnification probabilities derived from each sample is statistically significant, a large number of light curves must be generated for each such parameter combination. In this study, 35000 light curves with a uniform redshift distribution in the range $z_\mathrm{QSO}=0.13$--3.6 ($\Delta z_\mathrm{QSO} = 0.01$ between each bin) have been considered adequate. Simple tests show that this number results in $P_{min}$ derived from the final samples which are sufficiently stable for the present purpose, i.e. to establish the approximate border between rejected and allowed regions of the microlensing parameter space. 

\begin{table}[h]
 \caption[]{The set of discrete parameter values which define the grid of synthetic light curve samples used in this study.} 
\begin{flushleft}
\begin{tabular}{lllllll} 
\hline
$\Omega_\mathrm{compact}$: & 0.05 & 0.1 & 0.15 & 0.2 & 0.25 & 0.30\cr
$\log M_\mathrm{compact} \ (M_\odot)$:  &  -4 & -3 & -2 & -1 & 0 &\cr 
$\log R_\mathrm{QSO} \ (\mathrm{m})$: & 12 & 12.5 & 13 & 13.5 & 14 &\cr
\hline
\end{tabular}
\end{flushleft}
\end{table}

\section{Upper limits on the cosmological density of compact objects}
\subsection{Upper Limits on $\Omega_\mathrm{compact}$ inferred from the samples of Hawkins \& V\'eron (\cite{Hawkins & Véron})}
In HV93, two observational quasar samples are described: a FOCAP sample with a limiting magnitude of $m_\mathrm{B}=21$ and a bright sample with threshold $m_\mathrm{B}=18.5$. All objects in these samples have been monitored for a period of 10 years. Following S93, these two samples have been combined into one, consisting of 117 objects in the redshift range $z_\mathrm{QSO}=0.29$--3.23. 
In the context of deriving upper limits on $\Omega_\mathrm{compact}$ from light curve amplitudes, this HV93 sample is however not ideal, since the variability is not explicitly expressed as an amplitude. Instead, a related parameter $s$ is used, which may only approximately be transformed into an amplitude through the relation $\delta m\approx s/10$. Because of this complication, the constraints on $\Omega_\mathrm{compact}$ derived from the HV93 sample should be regarded as less certain than those inferred from the samples of H2000, and are only included here to facilitate a comparison of this study to that of S93. 

When calculating the synthetic $\delta m$ distribution, the time sampling must also be carefully considered. Even though the measurements of HV93 span 10 years, the $s$ parameter is only derived from the 7 years when more than one plate was taken. This implies that the amplitudes of the synthetic light curves should be derived from 7 data points with the same separation as in HV93 (Hawkins 2002, private communication), not from the 11 evenly spaced data points assumed in S93. The effect of using too many yearly data points is to increase the probability of observing large variations, thereby making the constraints inferred on $\Omega_\mathrm{compact}$ too strong. When calculating the amplitudes, S93 furthermore neglects the averaging over intrayear magnitude variations described in section 3.1, thereby inferring to strong constraints on scenarios in which the typical time scale of variations is smaller than one year.

In Fig.~\ref{HV93, 7dp} we display the limits on $\Omega_\mathrm{compact}$ inferred from the HV93 sample when the more realistic time sampling described above is used and assuming that $R_\mathrm{QSO}$ may be considered a free parameter. As seen, the efficiency of this method to impose upper limits on $\Omega_\mathrm{compact}$ is very sensitive to the value of $R_\mathrm{QSO}$ assumed. For $R_\mathrm{QSO} \le 1 \cdot 10^{13}$ m, strong constraints of $\Omega_\mathrm{compact} \le 0.05$--0.1 may be imposed on compact objects in the mass interval $M_\mathrm{compact}=10^{-3}\ M_\odot$--$10^{-2} \ M_\odot$. At $R_\mathrm{QSO}=3 \cdot 10^{13}$ m however, the upper limits are significantly weaker and at $R_\mathrm{QSO} = 10^{14}$ m no meaningful constraints may be imposed.
 
In Fig.~\ref{HV93, 7dp + VAR, z>1.5 RQSO-LQSO} we also indicate the constraints on $\Omega_\mathrm{compact}$ when the $L_\mathrm{QSO}$-$R_\mathrm{QSO}$ relation (\ref{eq22}) is used. The constraints are identical to those produced for $R_\mathrm{QSO} =10^{13}$ m, which is explained by the strong peak located at this value in the predicted $R_\mathrm{QSO}$-distribution. Fig.~\ref{HV93, 7dp + VAR, z>1.5 RQSO-LQSO, RQSOdists} illustrates the $R_\mathrm{QSO}$-distribution of the synthetic HV93 sample when $\Omega_\mathrm{compact}=0.3$ and $M_\mathrm{compact}=10^{-4} \ M_\odot$. The distributions for other parameter combinations only differ slightly due to the effect of amplification bias.
\begin{figure}[t]
\resizebox{\hsize}{!}{\includegraphics{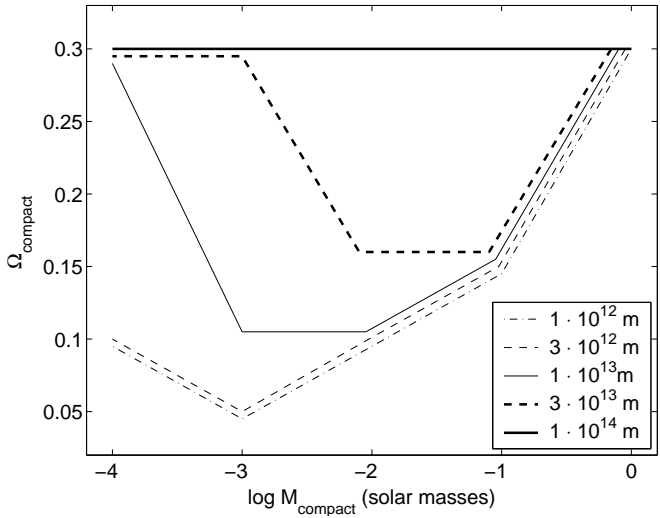}}
\caption[]{Upper limits on $\Omega_\mathrm{compact}$ for five different lens masses inferred from a comparison with the combined samples of HV93 when $R_\mathrm{QSO}$ is assumed unrelated to $L_\mathrm{QSO}$. The different lines represent $R_\mathrm{QSO}=10^{14}$ m (thick solid), $3\cdot10^{13}$ m (thick dashed), $10^{13}$ m (solid), $3\cdot10^{12}$ m (dashed) and $10^{12}$ m (dash-dotted). Slight offsets from the parameter values of Table 1, for which the formal upper limits are derived, have been introduced to prevent overlapping lines. The maximum upper limit of $\Omega_\mathrm{compact}=0.3$ is here set by the background cosmology ($\Omega_\mathrm{M}=0.3$, $\Omega_\Lambda=0.7$).}
\label{HV93, 7dp}
\end{figure}

\subsection{Upper limits on $\Omega_\mathrm{compact}$ inferred from the samples of Hawkins (\cite{Hawkins2000})}
We now shift our attention to the larger samples of H2000, which contain a total of 386 objects in the range $z_\mathrm{QSO}=0.136$--3.58, monitored for a period of 20 years. H2000 contains three subsamples: UVX, VAR and AMP. 

The UVX sample, which contains 184 objects in the redshift range $z=0.242$--2.21 is formed from a criterion of ultraviolet excess, $U-B < -0.2$ and a limiting magnitude of $m_\mathrm{B}=21.5$. 

The first variability-selected sample VAR, containing 298 objects in the redshift range $z=0.136$--3.58, has a limiting magnitude of $m_\mathrm{B}=21$ and an amplitude threshold of $\delta m >0.35$. The second variability-selected sample AMP, containing 66 objects in the redshift range $z=0.14$--2.07, has a limiting magnitude of $m_\mathrm{B}=21$ and an amplitude threshold of $\delta m >1.1$. 

The constraints on $\Omega_\mathrm{compact}$ that may be derived from a comparison of microlensing simulations to the statistical properties of these samples depend strongly on their selection criteria and redshift distributions. In order to find the most powerful upper limits, several combinations of the different subsamples have been tested. 
	
In the following, the light curves of the synthetic samples have been sampled at 21 evenly spaced yearly data points, each constituting an average of four intrayear magnification points, in fair agreement with the observational procedure (Hawkins 2002, private communication).

\begin{figure}[t]
\resizebox{\hsize}{!}{\includegraphics{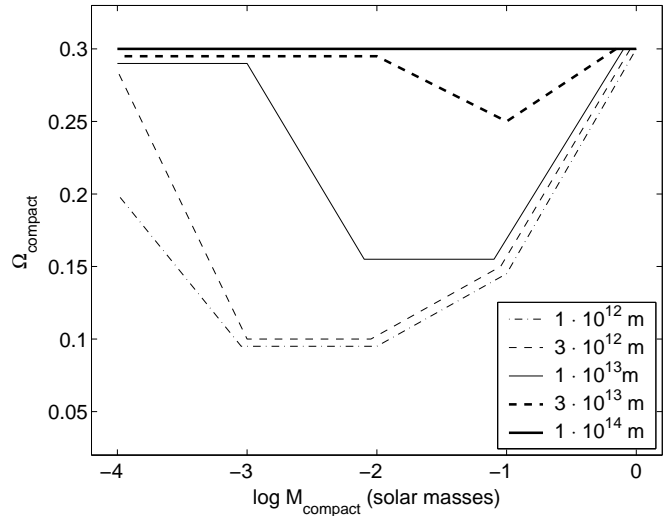}}
\caption[]{Same as Fig.~\ref{HV93, 7dp} for the UVX sample of H2000.}
\label{UVX}
\end{figure}

In Fig.~\ref{UVX}, we present the constraints on $\Omega_\mathrm{compact}$ inferred from the UVX data set when assuming $R_\mathrm{QSO}$ to be unrelated to $L_\mathrm{QSO}$. The constraints are weaker than those produced from a comparison to HV93 (Fig.~\ref{HV93, 7dp}).  The most important reason for this is the different observational selection criterion used, which prevents high redshift objects from entering the UVX sample. At $z>2.2$ the Lyman forest enters the U-band, rendering the quasars red and out of reach of the $U-B<-0.2$ criterion. 

It is reasonable to expect that a comparison to the variability-selected samples should result in more powerful upper limits on $\Omega_\mathrm{compact}$ than those inferred from UVX, since these samples may extend to much higher redshifts, and the probabilities of high amplitudes increase with redshift in the microlensing paradigm, but not - as shown in H2000 - in the observations. The main obstacle of this approach lies in the amplitude thresholds, which overestimates the true variation probability by rejecting low-amplitude objects.  If the amplitude threshold is too high, no meaningful constraints on $\Omega_\mathrm{compact}$ can be imposed.

\begin{figure}[t]
\resizebox{\hsize}{!}{\includegraphics{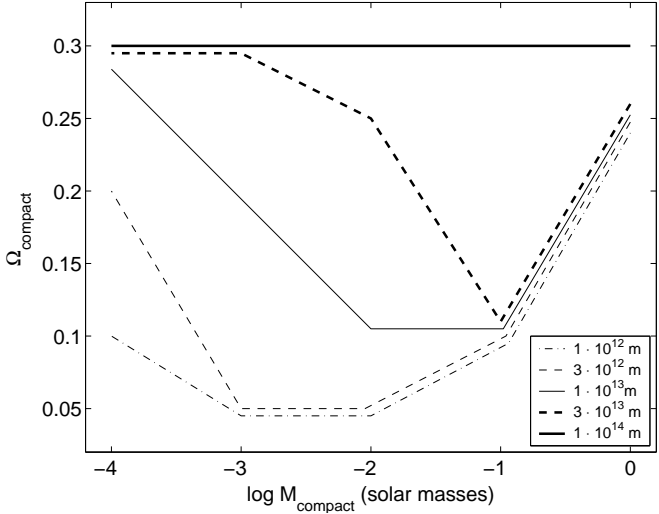}}
\caption[]{Same as Fig.~\ref{HV93, 7dp} for the UVX+VAR samples of H2000.}
\label{UVX+VAR}
\end{figure}

One may also combine UVX and variability selected samples to form a data set which both extends to high redshifts and is less likely to substantially overestimate the variation probability of quasars. In Fig.~\ref{UVX+VAR} we indicate the constraints on $\Omega_\mathrm{compact}$ inferred from the combined UVX+VAR sample when $R_\mathrm{QSO}$ is assumed unrelated to $L_\mathrm{QSO}$. Due to the extension to higher redshifts, the upper limits are significantly stronger than those inferred from UVX. The upper limits derived from the VAR sample alone are somewhat weaker than those presented in Fig.~\ref{UVX+VAR}. 

The AMP sample is not useful for these purposes, both because of its 
very high amplitude threshold and its limited redshift span. 

Since the probabilities of high amplitudes increase with redshift in the microlensing scheme, a possible way to improve the efficiency of this method even further may however be to conduct a comparison between a high-redshift subset of the VAR sample and the microlensing simulations.

H2000 shows that the higher redshift objects in the VAR sample display less variability than their low redshift counterparts. This is attributed to the fact that the higher redshift quasars also have higher luminosities and the hypothesis that $R_\mathrm{QSO}$ increases with $L_\mathrm{QSO}$. However, by stretching the assumption that these two parameters are unrelated, the fact that the high redshift quasars of VAR show smaller amplitudes than their low redshift counterparts should not hamper the exclusive use of high redshift objects in the comparison. Fig.~\ref{VAR, z>1.5} shows the substantially improved constraints when all objects with $z<1.5$ have been removed from the VAR sample. This leaves 144 objects, which still provides better statistics than the samples of HV93. The constraints inferred this way are very strong: $\Omega_\mathrm{compact}\le 0.05$ for $M_\mathrm{compact}=10^{-3}\ M_\odot-10^{-1} \ M_\odot$ and $\Omega_\mathrm{compact}\le 0.1$ for $M_\mathrm{compact}=1 \ M_\odot$ as long as $R_\mathrm{QSO}\le 10^{13}$ m. Even at $R_\mathrm{QSO}= 3\cdot 10^{13}$ m competitive constraints may be derived for $M_\mathrm{compact}=10^{-2}\ M_\odot-10^{-1} \ M_\odot$.

Even though it may be possible to construct a $L_\mathrm{QSO}$-$R_\mathrm{QSO}$ relation which mimics the inverse correlation between luminosity and amplitude seen in the observed samples, while allowing essentially no amplitude dependence on redshift for $z>0.5$, the $L_\mathrm{QSO}$-$R_\mathrm{QSO}$ relation explored here does not have this property. In Fig.~\ref{HV93, 7dp + VAR, z>1.5 RQSO-LQSO} we indicate the constraints on $\Omega_\mathrm{compact}$ derived from a comparison to the $z>1.5$ part of VAR in the case when the $L_\mathrm{QSO}$-$R_\mathrm{QSO}$ relation (\ref{eq22}) is used. In Fig.~\ref{HV93, 7dp + VAR, z>1.5 RQSO-LQSO, RQSOdists}, we show the corresponding $R_\mathrm{QSO}$-distribution of the sample generated when $\Omega_\mathrm{compact}=0.3$ and $M_\mathrm{compact}=10^{-4} \ M_\odot$. Despite the higher redshift, the typical $R_\mathrm{QSO}$ value actually lies somewhat lower than that predicted for the HV93 sample. The low-$R_\mathrm{QSO}$ tail not seen in the HV93 distribution is a result of the stronger amplification bias induced at high redshift, where the average intrinsic quasar luminosity lies further below the limiting luminosity of the sample. 

\section{Discussion}
As indicated in Fig.~\ref{VAR, z>1.5}, the method of using the long-term variability of quasars provides excellent possibilities to constrain $\Omega_\mathrm{compact}$, especially with the use of high-redshift quasar samples. For certain combinations of $M_\mathrm{compact}$ and $R_\mathrm{QSO}$, $P_\mathrm{min}$ falls far below the threshold value of 10\% even at the lowest values of $\Omega_\mathrm{compact}$ considered, indicating that the method could possibly be refined to constrain cosmological densities even below $\Omega_\mathrm{compact}=0.05$. Similarly, we note that it is not impossible that constraints on $M_\mathrm{compact}=10^{-5} \ M_\odot$ may also be imposed. Due to the very CPU-demanding simulations necessary at such low masses (many lenses contributing to the light curve), we have however not investigated this possibility here.

Despite the improvement of the upper limits imposed here compared to those derived in S93, several uncertainties still prevent us from drawing definite conclusions about $\Omega_\mathrm{compact}$ in the 
mass range explored.  

As already shown, the constraints are highly sensitive to the typical value of the $R_\mathrm{QSO}$ parameter. Even though the few available measurements of $R_\mathrm{QSO}$ indicate values inside the region for which powerful constraints may be derived, extrapolation of these results to the whole quasar population could be hazardous. 

Other parameter uncertainties also deserve consideration. The upper limits on $\Omega_\mathrm{compact}$ presented so far rely on the assumption that $\sigma_{\mathrm{v,compact}}=400$ km/s. This parameter essentially describes the preferred whereabouts of the compact objects. If the compact objects cluster only weakly, i.e. are mainly located in the field, a value of $\sigma_{\mathrm{v,compact}}=200$ km/s should be more appropriate. If they cluster more strongly, i.e. are mainly distributed inside rich galaxy clusters, much higher values could in principle be considered.
An estimate of the upper limit to this value may be obtained from model predictions by Sheth et al. (\cite {Sheth}) under the assumption that the lenses dynamically behave similar to cold dark matter 
particles.  Based on the values they derive for different spatial scales we 
choose as an upper limit $\sigma_{\mathrm{v,compact}} = 600$ km/s.
\begin{figure}[t]
\resizebox{\hsize}{!}{\includegraphics{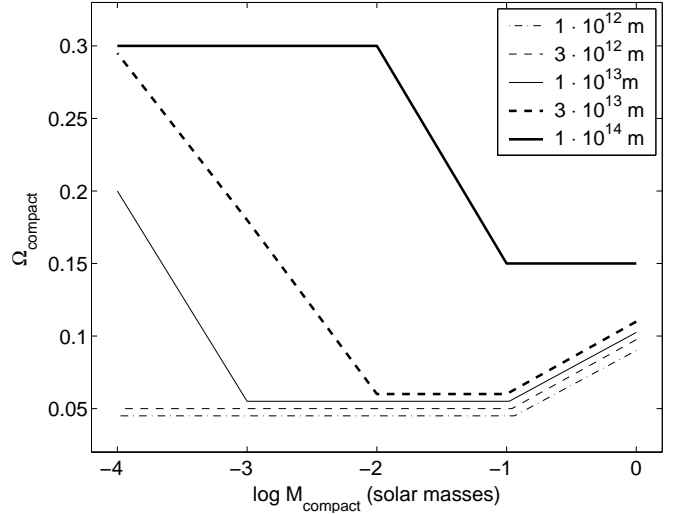}}
\caption[]{Same as Fig.~\ref{HV93, 7dp} for the $z>1.5$ part of the VAR sample of H2000.}
\label{VAR, z>1.5}
\end{figure}
\begin{figure}[t]
\resizebox{\hsize}{!}{\includegraphics{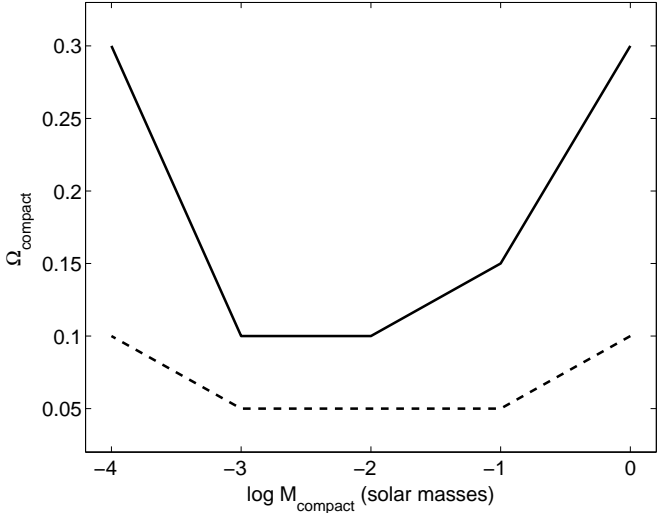}}
\caption[]{Upper limits on $\Omega_\mathrm{compact}$ inferred for different $M_\mathrm{compact}$ from the HV93 sample (thick solid) and the $z>1.5$ part of VAR (thick dashed) when the Czerny et al. (\cite{Czerny et al.}) $L_\mathrm{QSO}$-$R_\mathrm{QSO}$ relation is applied.}
\label{HV93, 7dp + VAR, z>1.5 RQSO-LQSO}
\end{figure}

In the context of microlensing, the effect of varying $\sigma_{\mathrm{v,compact}}$ is to alter the characteristic time scale of source crossing and the width of the corresponding magnification peak. Higher velocities imply shorter time scales and, in principle, an increased probability of detecting high amplitudes since more compact object may pass the source during the time span of the observational programme. The impact of this effect is however weakened by the finite sampling rate of the light curves. 

By expanding the grid of synthetic samples defined by Table 1 to other $\sigma_{\mathrm{v,compact}}$ at $R_\mathrm{QSO}=10^{13}$ m, we have investigated in what way the upper limits on $\Omega_\mathrm{compact}$ inferred from the $z>1.5$ subset of the VAR sample are affected by $\sigma_{\mathrm{v,compact}}$ variations in the range $\sigma_{\mathrm{v,compact}}=200$--600 km/s. The effect turns out to be very undramatic and only relaxes our upper limits from $\Omega_\mathrm{compact}=0.1$ to 0.15 at 
$M_\mathrm{compact}=1 \ M_\odot$ in the case of $\sigma_{\mathrm{v,compact}}=200$ km/s. All other constraints at this source size are unaffected.

Additional uncertainties stem from approximations inherent in the microlensing model used. The model assumes no shear and a circular source with uniform brightness. The inclusion of shear terms is expected to lower $\Delta m$ (Refsdal \& Stabell \cite{Refsdal & Stabell}), thereby making the constraints on $\Omega_\mathrm{compact}$ weaker. Due to the lack of analytical approximations for $\mu(y,R)$ in the case of both non-zero shear and more realistic (e.g. Gaussian) source brightness profiles, the quantitative impact of such features can however presently only be tested with more time-consuming algorithms like backwards ray-tracing. The model also assumes all compact objects to be randomly distributed, i.e. that correlations between lenses are unimportant. However, for the most extreme combination of lens parameters considered in this study, the  simulations include compact objects initially as far away as two light years from each other in the direction perpendicular to the line-of-sight. This means that compact objects which are bound together on scales smaller than this (e.g. in clusters or binary systems) will have correlated positions in the lens plane. Simple tests show that correlations of this kind generally increase the probability of high amplitudes, which indicates that the upper limits derived from random lens distributions should be conservative.

\begin{figure}[h]
\resizebox{\hsize}{!}{\includegraphics{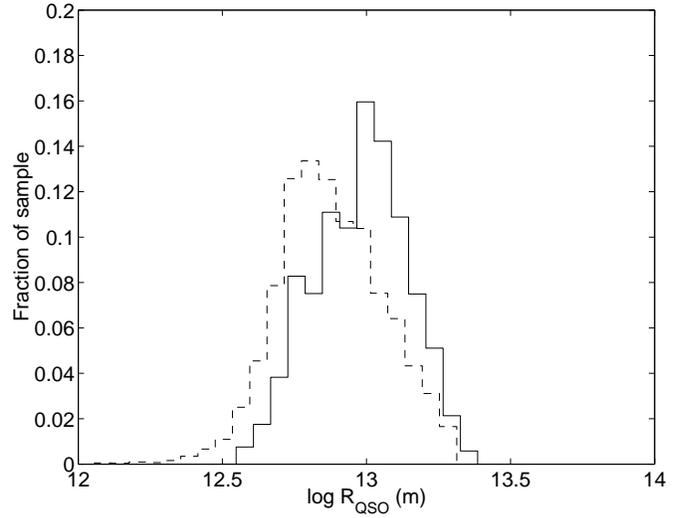}}
\caption[]{Examples of the $R_\mathrm{QSO}$ distributions of the synthetic samples generated to match the HV93 sample (solid) and the $z>1.5$ subset of VAR (dashed), when the Czerny et al. (\cite{Czerny et al.}) $L_\mathrm{QSO}$-$R_\mathrm{QSO}$ relation is applied. In both cases we assume $\Omega_\mathrm{compact}=0.3$, $M_\mathrm{compact}=10^{-4} \ M_\odot$ and use identical bin widths when plotting the distributions.}
\label{HV93, 7dp + VAR, z>1.5 RQSO-LQSO, RQSOdists}
\end{figure}

\section{Conclusions} 
In this paper, the analysis of the upper limits on the cosmological density of dark matter in the form of compact objects inferred from the long-term variability of quasars has been improved compared to that of S93 in several ways: 

\begin{itemize} 
\item By generalization to the $\Omega_\mathrm{M}=0.3$, $\Omega_\Lambda=0.7$ universe; 
\item By implementation of a magnification formula more suitable for microlensing of large sources; 
\item By implementation of an improved method of generating synthetic quasar samples which matches the $z_\mathrm{QSO}$ and $m_\mathrm{B}$ distributions of the observational data sets; 
\item By considering uncertainties in the typical value of $R_\mathrm{QSO}$ and $\sigma_\mathrm{v,compact}$;
\item By taking the intrayear magnitude averaging of the observational procedure into account when deriving the light curve amplitudes; 
\item By comparing results from numerical simulations to larger observational 
samples with longer time span of quasar monitoring.
\end{itemize}

We show that by using a high redshift subset of variability-selected samples, previous limits may be substantially improved. We infer $\Omega_\mathrm{compact}\le 0.05$ for $M_\mathrm{compact}=10^{- 
3}\ M_\odot$--$10^{-1}\ M_\odot$ as long as typically $R_\mathrm{QSO}\le 1\cdot 10^{13}$ m at $z>1.5$. The few existing measurements of $R_\mathrm{QSO}$ are however not sufficient to ensure that this condition is fulfilled. At $R_\mathrm{QSO}=3\cdot 10^{13}$ m the $\Omega_\mathrm{compact}\le 0.05$ constraint only holds for lenses in the mass range $M_\mathrm{compact}=10^{- 
2}\ M_\odot$--$10^{-1}\ M_\odot$, and at $R_\mathrm{QSO}=10^{14}$ m the only meaningful constraints are $\Omega_\mathrm{compact}\le 0.15$ for $M_\mathrm{compact}=10^{-1}\ M_\odot$--$1 \ M_\odot$. 

When the strong $R_\mathrm{QSO}$-dependence of the upper limits is compounded with additional uncertainties discussed under Section 5, we are forced to conclude that the method of using the long-term variability of quasars to place upper limits on $\Omega_\mathrm{compact}$ cannot be used to reliably rule out any cosmologically significant populations of compact objects in the stellar to planetary mass range at the present time. This situation may of course change as more and better measurements of $R_\mathrm{QSO}$ become available.

\begin{acknowledgements}  
We greatfully thank Patrik Gullerstr\"om for his contributions to the programming of the microlensing code. 
\end{acknowledgements}

\end{document}